\documentclass[preprint,5p,times,twocolumn]{elsarticle}

\usepackage{amssymb,amsmath}
\usepackage{hyperref}

\journal{Computational Materials Science}

\begin{document}

\begin{frontmatter}



\title{\textit{Ab initio} calculation of transport and optical properties of aluminum: influence of simulation parameters}



\author[JIHT,MIPT]{D.V.~Knyazev\corref{cor1}}
\ead{d.v.knyazev@yandex.ru}

\author[JIHT,MIPT]{P.R.~Levashov}

\cortext[cor1]{Corresponding author}

\address[JIHT]{Joint Institute for High Temperatures RAS, Izhorskaya 13 bldg. 2, Moscow 125412, Russia}
\address[MIPT]{Moscow Institute of Physics and Technology (State University), Institutskii per. 9, Dolgoprudny, Moscow Region, 141700, Russia}

\begin{abstract}
This work is devoted to the \textit{ab initio} calculation of transport and optical properties of aluminum. The calculation is based on the quantum molecular dynamics simulation, density functional theory and the Kubo-Greenwood formula. Mainly the calculations are performed for liquid aluminum at near-normal densities for the temperatures from melting up to 20000~K. The results on dynamic electrical conductivity, static electrical conductivity and thermal conductivity are obtained and compared with available reference and experimental data and the calculations of other authors. The influence of the technical parameters on the results is investigated in detail. The error of static electrical conductivity calculation is estimated to be about 20\%; more accurate results require bigger number of atoms.
\end{abstract}

\begin{keyword}
quantum molecular dynamics \sep density functional theory \sep Kubo-Greenwood formula \sep transport and optical properties \sep aluminum
\end{keyword}

\end{frontmatter}


\section{INTRODUCTION}
\label{Sec:Introduction}

The knowledge of the transport and optical properties of matter in the warm dense matter regime is crucial for many modern fundamental and applied researches: femtosecond laser heating \cite{Veysman:JPB:2008}, experiments with exploding wires \cite{DeSilva:PRE:1998}, investigation of the metal-nonmetal transition \cite{Korobenko:PRB:2012}.

\textit{Ab initio} calculations of the transport and optical properties based on the quantum molecular dynamics (QMD) simulation, finite temperature density functional theory (FT-DFT) and the Kubo-Greenwood formula are widely used nowadays. The computations are performed for the substances of different types: liquid metals \cite{Desjarlais:PRE:2002}, hydrogen \cite{Collins:PRB:2001}, molten salts \cite{Silvestrelli:PRB:1996}, silica \cite{Laudernet:PRB:2004} and plastics \cite{Lambert:PRE:2012}.

The paper \cite{Silvestrelli:PRB:1999} is one of the first works that applies the method specified to the calculation of transport and optical properties of aluminum. The works \cite{Desjarlais:PRE:2002, Mazevet:PRE:2005, Recoules:PRE:2002} report optical properties and the static electrical conductivity of aluminum in the region on the phase diagram available with exploding wire experiments. The obtained results on the static electrical conductivity are compared with the exploding wire experiments \cite{DeSilva:PRE:1998,  Recoules:PRE:2002}. Other papers \cite{Recoules:PRB:2005, Alemany:PRB:2004, Knider:JPCM:2007} provide the comparison with slow thermophysical experiments on the optical properties, static electrical conductivity and thermal conductivity.

The number of atoms in the supercell is an important technical parameter during the calculation according to the method discussed. The number of atoms used in the papers on aluminum mentioned above is usually not larger than 108. Our previous paper \cite{Povarnitsyn:CPP:2012} reports results on the optical properties of aluminum calculated with 108 atoms in the supercell. Only the work \cite{Alemany:PRB:2004} demonstrates the computation with 205 atoms.

On the other hand, the most recent works \cite{Lambert:PhysPlasmas:2011} and \cite{Pozzo:PRB:2011} show significant influence of the size effects on the transport and optical properties. The supercell should contain 1000 --- 2000 atoms for convergence with the number of atoms to be achieved. The calculations in the papers \cite{Lambert:PhysPlasmas:2011} and \cite{Pozzo:PRB:2011} are performed for hydrogen and sodium, respectively. The recent paper \cite{Alfe:PRB:2012} shows that the increasing of the number of atoms up to 1024 is necessary to achieve well-converged results for solid iron.

In this paper we perform calculations for aluminum, taking into account three electrons per atom. Mainly we use 256 atoms in the supercell. The influence of the main technical parameters on the results is discussed. We vary the technical parameters and estimate the error of the static electrical conductivity calculation.
 
The transport and optical properties of aluminum were mainly calculated for the liquid phase at near-normal densities for the temperatures from melting up to 20000~K. The results are compared with available experimental and reference data and calculations by other authors.

Our paper is organized as follows. The method of calculation is described in section~\ref{Sec:Method}. Here the main technical parameters are introduced. The section~\ref{Sec:Technique} gives the numerical values of the technical parameters. The choice of these parameters is explained, the convergence according to these parameters is investigated, the error of the static electrical conductivity is estimated. The section~\ref{Sec:Onsager} contains some technical information on the expression for the calculation of the dynamic Onsager coefficients. The main results on the optical properties, static electrical and thermal conductivity are described in section~\ref{Sec:Results}.

\section{METHOD OF CALCULATION}
\label{Sec:Method}

The calculation consists of three main parts: quantum molecular dynamics simulation, precise band structure resolution and calculation of transport and optical properties via the Kubo-Greenwood formula.

Initially atoms are placed to the supercell with periodic boundary conditions. Then during the QMD simulation at given density and temperature the ionic trajectories are calculated. The independent ionic configurations are selected from the equilibrium section of the QMD simulation for the further calculation of optical properties and thermal conductivity.

Precise band structure calculation is performed for the selected ionic configurations. As well as during the QMD simulation electronic structure is calculated via the density functional theory. Technical parameters, used during the precise band structure resolution lead to more accurate results than during the QMD simulation. The obtained wave functions are used to calculate the matrix elements of the nabla operator.

The real part of the dynamic electrical conductivity is calculated via the Kubo-Greenwood formula (\ref{Eq:KuboGreenwoodsigma}) using the matrix elements, energy eigenvalues and Fermi-weights obtained during the precise resolution of the band structure. The values of the dynamic electrical conductivity obtained for the different ionic configurations are averaged. The imaginary part of the dynamic electrical conductivity is reconstructed via the Kramers-Kronig transformation (\ref{Eq:Kramers}). If the real and imaginary parts of the dynamic electrical conductivity are known the optical properties may be calculated.

The dynamic Onsager coefficients are calculated via the Kubo-Greenwood formula (\ref{Eq:DynamicOnsager}). The values of the Onsager coefficients obtained for the different ionic configurations are averaged. The static Onsager coefficients are obtained by extrapolation of the dynamic to the zero frequency. Thermal conductivity is expressed via the static Onsager coefficients (\ref{Eq:KExact}).

\subsection{Quantum molecular dynamics simulation}

During the QMD simulation the atoms at given density are placed to the cubic supercell with periodic boundary conditions.

At each step of the QMD simulation the electronic structure is calculated within the framework of density functional theory (DFT). The calculation is performed in the Born-Oppenheimer approximation: due to the great difference in masses between electrons and ions, the electrons immediately adjust to the current spatial positions of ions.

The electronic structure is calculated by solving the Kohn-Sham equations at the finite temperature \cite{Martin:2004}:

\begin{multline}
\left[-\frac{\hbar^2}{2m}\nabla^2+v_\mathrm{ext}(\mathbf{r})+\int\frac{\rho(\mathbf{r}^\prime)}{\left|\mathbf{r}-\mathbf{r}^\prime\right|}d\mathbf{r}^\prime+V_\mathrm{xc}(\mathbf{r})\right]\Psi_i(\mathbf{r})=\\=\epsilon_i\Psi_i(\mathbf{r}),
\label{Eq:KohnSham}
\end{multline}
\begin{gather}
\rho(\mathbf{r})=\sum_i f(\epsilon_i)\left|\Psi_i(\mathbf{r})\right|^2,
\label{Eq:DensityWF}
\\
f(\epsilon_i)=\frac{1}{\exp{\frac{\epsilon_i-\mu}{kT}}+1},
\label{Eq:FermiDistribution}
\\
\sum_i f(\epsilon_i)=N.
\label{Eq:MuDetermination}
\end{gather}
Here $\Psi_i(\mathbf{r})$ are the Kohn-Sham wave functions, $\epsilon_i$~--- the corresponding energy eigenvalues, $f(\epsilon_i)$~--- the corresponding Fermi-weights, $\rho(\mathbf{r})$~--- the electron density, $\mu$~--- the chemical potential, $N$~--- the number of electrons included in calculation, $\hbar$~--- the Planck constant, $m$ and $e$~--- the electron mass and charge respectively.

$V_\mathrm{xc}(\mathbf{r})$ is the exchange-correlation potential defined as a variation of exchange-correlation functional

\begin{equation}
V_\mathrm{xc}(\mathbf{r})=\frac{\delta E_\mathrm{xc}\left[\rho(\mathbf{r})\right]}{\delta\rho(\mathbf{r})},
\end{equation}
$v_\mathrm{ext}(\mathbf{r})$ is the ion potential. The ions are treated within the pseudopotential approach, which significantly reduces computational effort. The choice of the pseudopotential and exchange-correlation functional is discussed below (see~\ref{Subsec:Technique:Pseudopotential}).
  
The electron temperature is introduced by the parameter $T$ in the Fermi-Dirac distribution (\ref{Eq:FermiDistribution}).

The solution to the Kohn-Sham equations (\ref{Eq:KohnSham})--(\ref{Eq:MuDetermination}) for a supercell with periodic boundary conditions is the energy eigenvalue $\epsilon_{i,\mathbf{k}}$ and the corresponding wave function $\Psi_{i,\mathbf{k}}(\mathbf{r})$ which is represented as the set of plane wave orbitals:

\begin{equation}
\Psi_{i,\mathbf{k}}(\mathbf{r})=\frac{1}{\sqrt{\Omega}}\sum_{\mathbf{G}}e^{i(\mathbf{k}+\mathbf{G})\mathbf{r}}c_{i,\mathbf{k},\mathbf{G}}\,.
\label{Eq:Planewaves}
\end{equation}
Here $\Omega$ is the volume of the supercell, $c_{i,\mathbf{k},\mathbf{G}}$ are the coefficients of the expansion over plane waves, \textbf{G} is so-called \textbf{G}-vector, its coordinates are multiples of reciprocal lattice basis vectors. The number of the \textbf{G}-vectors is controlled by the energy cut-off $E_\mathrm{cut}$:

\begin{equation}
\frac{\hbar^2}{2m}\left|\mathbf{G}+\mathbf{k}\right|^2<E_\mathrm{cut}.
\end{equation} 
The energy cut-off $E_\mathrm{cut}$ is the important technical parameter; its choice is discussed below (see~\ref{Subsec:Technique:Encut}).

\textbf{k}-vectors in the expansion (\ref{Eq:Planewaves}) are called \textbf{k}-points in the Brillouin zone. The number of \textbf{k}-points is also an important technical parameter, we briefly discuss its influence on the results in \ref{Sec:Appendix}.

Each solution to the Kohn-Sham equations (\ref{Eq:KohnSham})--(\ref{Eq:MuDetermination}) in the supercell with periodic boundary conditions is called a \textit{band} in this paper. The number of bands $N_\mathrm{bands}$ included to the solving of the Kohn-Sham equations is a technical parameter, discussed below (see~\ref{Subsec:Technique:Nbands}).

The forces acting on the ions are calculated during the QMD simulations via the Hellmann-Feynman theorem. Ions are treated classically. The Newton equations acting on the ions are solved using the Verlet algorithm. The temperature of ions is maintained via the Nose-Hoover thermostat.

The equilibrium section of the QMD simulation is used to choose the ionic configurations for the further precise resolution of the band structure. The choice of ionic configurations is discussed below (see~\ref{Subsec:Technique:Configurations}).

\subsection{Precise resolution of the band structure}

The ionic configurations selected during the QMD simulation are used to calculate the band structure. For these configurations the Kohn-Sham equations (\ref{Eq:KohnSham})--(\ref{Eq:MuDetermination}) are solved one more time, but with the larger energy cut-off, number of bands, number of \textbf{k}-points in the Brillouin zone. The ions are not moved. This stage is referred to as the \textit{precise resolution of the band structure} in this work.

The technical parameters used in the precise resolution of the band structure are discussed in section~\ref{Sec:Technique}.

The Kohn-Sham wave functions $\Psi_i(\mathbf{r})$ are necessary to calculate the matrix elements of the nabla operator $\left\langle\Psi_i\left|\nabla_{\alpha}\right|\Psi_j\right\rangle$. Some technical details about the calculation of the matrix elements are discussed below (see~\ref{Subsec:Technique:Pseudopotential}).

\subsection{Calculation of optical properties}

The complex dynamic electrical conductivity $\sigma(\omega)=\sigma_1(\omega)+i\sigma_2(\omega)$ is the complex coefficient between the current density $\mathbf{j}_\omega$ and applied electric field $\mathbf{E}_\omega$ of the frequency $\omega$:
\begin{equation}
\mathbf{j}_\omega=(\sigma_1(\omega)+i\sigma_2(\omega))\mathbf{E}_\omega.
\label{Eq:SigmaDefinition}
\end{equation}
 
The real part of the dynamic electrical conductivity (also referred to as simply dynamic electrical conductivity in this work) $\sigma_1(\omega)$ is connected with the energy absorbed by the electrons. It is calculated via the Kubo-Greenwood formula for each ionic configuration:

\begin{multline}
\sigma_1(\omega)=\frac{2\pi e^2\hbar^2}{3m^2\omega\Omega}\times\\
\sum_{i,j,\alpha,\mathbf{k}}W(\mathbf{k})\left|\left\langle\Psi_{i,\mathbf{k}}\left|\nabla_\alpha\right|\Psi_{j,\mathbf{k}}\right\rangle\right|^2\times\\
\left[f(\epsilon_{i,\mathbf{k}})-f(\epsilon_{j,\mathbf{k}})\right]\delta(\epsilon_{j,\mathbf{k}}-\epsilon_{i,\mathbf{k}}-\hbar\omega).
\label{Eq:KuboGreenwoodsigma}
\end{multline}
Here sum is over all $\mathbf k$-points within the Brillouin zone (by $\mathbf k$), all bands participating in the calculation (by $i$ and $j$), and three spatial dimensions (by $\alpha$). $W(\mathbf{k})$ is the weight of the particular \textbf{k}-point in the Brillouin zone. A simple, short and intuitively clear derivation of the Kubo-Greenwood formula for the dynamic electrical conductivity may be found in \cite{Moseley:AJP:1978}.

For the practical calculation the $\delta$-function in the Kubo-Greenwood formula (\ref{Eq:KuboGreenwoodsigma}) should be broadened by the Gaussian function:

\begin{multline}
\delta(\epsilon_{j,\mathbf{k}}-\epsilon_{i,\mathbf{k}}-\hbar\omega)\stackrel{}{\rightarrow}\\
\stackrel{}{\rightarrow}\frac{1}{\sqrt{2\pi}\Delta E}\exp\left(-\frac{(\epsilon_{j,\mathbf{k}}-\epsilon_{i,\mathbf{k}}-\hbar\omega)^2}{2(\Delta E)^2}\right),
\label{Eq:Deltabroadening}	
\end{multline}
where $\Delta E$ is the broadening of the $\delta$-function. The broadening $\Delta E$ is an important technical parameter, the choice of its value is described below (see~\ref{Subsec:Technique:Broadening}).

The values of the dynamic electrical conductivity obtained for the different ionic configurations are averaged.

If the real part of the dynamic electrical conductivity is known, the imaginary part may be reconstructed via the Kramers-Kronig transformation \cite{Collins:PRB:2001}:

\begin{equation}
\sigma_2\left(\omega\right)=-\frac{2}{\pi}P\int_0^\infty\frac{\sigma_1\left(\nu\right)\omega}{\nu^2-\omega^2}d\nu,
\label{Eq:Kramers}
\end{equation}
where $P$ stands for the principle value of the integral.

If the complex dynamic electrical conductivity is known, the following optical properties may be calculated:

complex dielectric function $\varepsilon(\omega)=\varepsilon_1(\omega)+i\varepsilon_2(\omega)$:
\begin{equation}
\varepsilon_1\left(\omega\right)=1-\frac{\sigma_2\left(\omega\right)}{\omega\varepsilon_0};~~~ 
\varepsilon_2\left(\omega\right)=\frac{\sigma_1\left(\omega\right)}{\omega\varepsilon_0},
\label{Eq:Epsilon}
\end{equation}
where $\varepsilon_0$ is the dielectric permittivity of vacuum (SI units);

complex index of refraction $n(\omega)+ik(\omega)$:

\begin{align}
n(\omega)=\frac{1}{\sqrt{2}}\sqrt{\left|\varepsilon(\omega)\right|+\varepsilon_1(\omega)};\\
k(\omega)=\frac{1}{\sqrt{2}}\sqrt{\left|\varepsilon(\omega)\right|-\varepsilon_1(\omega)};
\end{align}

reflectivity (at the normal incidence of laser radiation):

\begin{equation}
r(\omega)=\frac{\left[1-n(\omega)\right]^2+k(\omega)^2}{\left[1+n(\omega)\right]^2+k(\omega)^2};
\end{equation}

absorption coefficient:

\begin{equation}
\alpha(\omega)=2k(\omega)\frac{\omega}{c}.
\end{equation}

\subsection{Calculation of thermal conductivity}

The Onsager coefficients $\mathcal{L}_{mn},~m,n=1,2$ (referred to as the \textit{static Onsager coefficients} in this work) are introduced as the coefficients that connect the electric current density $\mathbf{j}$ and heat current density $\mathbf{j}_q$ with the applied constant in time electric field $\mathbf{E}$ and temperature gradient $\nabla T$:

\begin{eqnarray}
\mathbf{j}=\frac{1}{e}\left(e\mathcal{L}_{11}\mathbf{E}-\frac{\mathcal{L}_{12}\nabla T}{T}\right),
\label{Eq:OnsagerDefinition1}
\\
\mathbf{j}_q=\frac{1}{e^2}\left(e\mathcal{L}_{21}\mathbf{E}-\frac{\mathcal{L}_{22}\nabla T}{T}\right).
\label{Eq:OnsagerDefinition2}
\end{eqnarray}

The Onsager coefficient $\mathcal{L}_{11}$ is the static electrical conductivity $\sigma_{1DC}$. The Onsager reciprocal relation is valid for the static coefficients $\mathcal{L}_{12}$ and $\mathcal{L}_{21}$ \cite{Ziman:1960}:

\begin{equation}
\mathcal{L}_{12}=\mathcal{L}_{21}
\end{equation}

The typical experiment on the thermal conductivity measurement is performed at the zero electric current density. Under this condition the thermal conductivity $K$ is defined as the factor between the heat current density and the applied temperature gradient:

\begin{equation}
\mathbf{j}_q=-K\nabla T.
\label{Eq:KDefinition}
\end{equation}
Setting to zero electric current density in (\ref{Eq:OnsagerDefinition1}) thermal conductivity may be expressed from (\ref{Eq:OnsagerDefinition2}) and (\ref{Eq:KDefinition}):

\begin{equation}
K=\frac{1}{e^2T}\left(\mathcal{L}_{22}-\frac{\mathcal{L}_{12}\mathcal{L}_{21}}{\mathcal{L}_{11}}\right).
\label{Eq:KExact}
\end{equation}

It may be shown \cite{Ashcroft:1976}, that at relatively low temperatures the second term in expression (\ref{Eq:KExact}) (referred to as the \textit{thermoelectric term} in this work) gives negligible contribution to the thermal conductivity and approximate expression is valid:

\begin{equation}
K\approx\frac{\mathcal{L}_{22}}{e^2T}.
\label{Eq:KApprox}
\end{equation}
In this work we calculate thermal conductivity via the exact expression (\ref{Eq:KExact}) and verify, whether approximate expression (\ref{Eq:KApprox}) is valid (see~\ref{Subsubsec:Results:Static:Normisochor}).

In this work we use the Kubo-Greenwood formula to calculate the dynamic Onsager coefficients $\mathcal{L}_{mn}(\omega)$:

\begin{multline}
\mathcal{L}_{mn}(\omega)=(-1)^{m+n}\frac{2\pi e^2\hbar^2}{3m^2\omega\Omega}\times\\
\sum_{i,j,\alpha,\mathbf{k}}W(\mathbf{k})\left(\frac{\epsilon_{i,\mathbf{k}}+\epsilon_{j,\mathbf{k}}}{2}-\mu\right)^{m+n-2}\left|\left\langle\Psi_{i,\mathbf{k}}\left|\nabla_\alpha\right|\Psi_{j,\mathbf{k}}\right\rangle\right|^2\\\times
\left[f(\epsilon_{i,\mathbf{k}})-f(\epsilon_{j,\mathbf{k}})\right]\delta(\epsilon_{j,\mathbf{k}}-\epsilon_{i,\mathbf{k}}-\hbar\omega).
\label{Eq:DynamicOnsager}
\end{multline}

The $\mathcal{L}_{11}(\omega)$ Onsager coefficient is the real part of the dynamic electrical conductivity and equation (\ref{Eq:DynamicOnsager}) reduces to equation (\ref{Eq:KuboGreenwoodsigma}). The other dynamic Onsager coefficients do not have direct physical meaning in this work and are used only to get the static Onsager coefficients by the extrapolation to the zero frequency. The static electrical conductivity is obtained by the extrapolation of $\mathcal{L}_{11}(\omega)=\sigma_1(\omega)$ to the zero frequency. The procedure of the extrapolation is described below (see~\ref{Subsec:Technique:Extrapolation}).

The equation (\ref{Eq:DynamicOnsager}) gives the equal values of $\mathcal{L}_{12}(\omega)$ and $\mathcal{L}_{21}(\omega)$ at non-zero frequencies. Some technical details connected with this fact are discussed below (section~\ref{Sec:Onsager}).

The values of the dynamic Onsager coefficients obtained for the different ionic configurations are averaged.

The derivation of the equation (\ref{Eq:DynamicOnsager}) may be found in \cite{Holst:PRB:2011}.

Finally, the experimentally discovered Wiedemann-Franz law should be mentioned. According to it, the ratio of the thermal conductivity $K$ to the product of static electrical conductivity and temperature $\sigma_1T$ is constant:

\begin{equation}
\frac{K(T)}{\sigma_1(T)\cdot T}=L=\frac{\pi^2}{3}\frac{k^2}{e^2},
\label{Eq:WiedemannFranz}
\end{equation}
where $k$ is the Boltzmann constant. The mentioned ratio $L$ is called the Lorenz number. Its value $L=\frac{\pi^2}{3}\frac{k^2}{e^2}=2.44\cdot 10^{-8}$~W$\cdot\Omega\cdot$K$^{-2}$ mentioned in equation (\ref{Eq:WiedemannFranz}) is called the ideal value. In the work \cite{Chester:PPS:1961} the Wiedemann-Franz law and the ideal value of the Lorenz number are explained theoretically under rather general assumptions. In this work we calculate the Lorenz number and thus verify the Wiedemann-Franz law.

In this work the QMD and band structure calculations are performed using the Vienna \textit{ab initio} simulation package (VASP) \cite{Kresse:PRB:1993, Kresse:PRB:1994, Kresse:PRB:1996}. We created a special parallel program module that computes the dynamic Onsager coefficients, extrapolates them to zero frequency and calculates the electrical and thermal conductivity. This module uses information on the wave functions or matrix elements from the VASP output (see~\ref{Subsec:Technique:Pseudopotential}).

\section{TECHNICAL DETAILS}
\label{Sec:Technique}

This section is devoted to the dependence of the dynamic and static electrical conductivity on the technical parameters. The choice of these parameters will be demonstrated for liquid aluminum at temperature $T=1273$~K and density $\rho=2.249$ g/cm$^3$.

In this work we do not vary the number of \textbf{k}-points and use only $\Gamma$-point in all calculations (see, however, \ref{Sec:Appendix}). In earlier work \cite{Desjarlais:PRE:2002} it was shown that for aluminum plasma the increase of the number of \textbf{k}-points gave insignificant effect on the corresponding electrical conductivity values. Nevertheless, we plan to clarify this question in our future work in detail.

Currently the calculation for this $(\rho, T)$ point is performed with the following technical parameters: 256~atoms, 1500~steps of QMD simulation, the step of the QMD simulation is 2~fs. During the QMD simulation ultrasoft pseudopotential of Vanderbilt \cite{Vanderbilt:PRB:1990} (US-PP) and local density approximation (LDA) for exchange-correlation functional are used. QMD simulation is performed with only 1 \textbf{k}-point in the Brillouin zone ($\Gamma$-point), energy cut-off is $E_\mathrm{cut}=100$~eV, the number of bands is 500. Ionic configurations for the calculation of transport and optical properties are taken every 100 steps. The precise resolution of band structure is performed for each of these ionic configurations using the same pair US-PP, LDA as during the QMD simulation. Band structure is calculated with only 1 \textbf{k}-point in the Brillouin zone ($\Gamma$-point), energy cut-off is $E_\mathrm{cut}=200$~eV, the number of bands is 1300. The dynamic electrical conductivity and the dynamic Onsager coefficients are calculated for the frequencies from 0.005~eV up to 10~eV with the step 0.005~eV. The broadening of the $\delta$-function in the Kubo-Greenwood formula is $\Delta E=0.1$~eV. The static Onsager coefficients are obtained by the linear extrapolation of the dynamic values to zero frequency.

The dependence of the results on the technical parameters is examined as follows: one parameter is varied while others are kept fixed at the values mentioned above. Then the obtained information is used to estimate the error of static electrical conductivity calculation.

\subsection{The ionic configurations}
\label{Subsec:Technique:Configurations}

\begin{figure}
	\includegraphics[width=0.95\columnwidth]{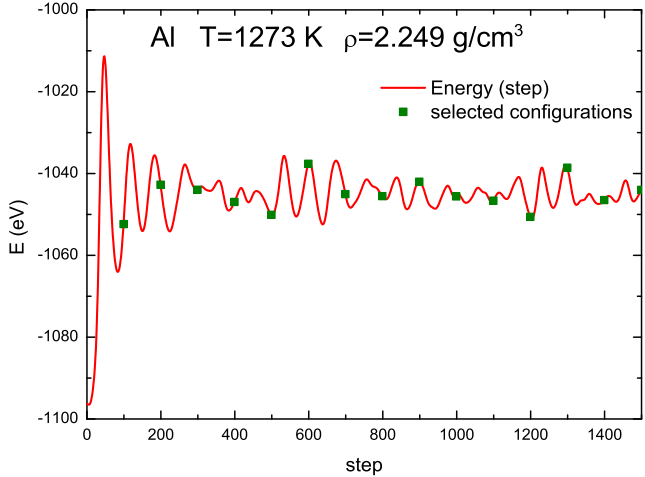}
	\caption{(Color online) The dependence of the total energy of electrons and ions on the step number. Solid line is the dependence of energy on the step number; square points are configurations selected for the further precise resolution of the band structure.}
	\label{Fig:QMDenergy}
\end{figure}

The typical dependence of the total energy of electrons and ions on the step number is shown in Fig.~\ref{Fig:QMDenergy}. At the start of simulation the atoms form the fcc lattice. The simulation is performed for 1500 steps, each step of simulation corresponds to 2~fs. Thus we observe the behavior of the system for 3~ps. During about 100 initial steps the lattice breaks down and energy comes to its equilibrium value. Then at the equilibrium section energy fluctuates around its average value. The configurations for the calculation of dynamic electrical conductivity are taken from this equilibrium section. The distance between neighbouring configurations is 100 steps.

To examine the convergence with the number of configurations $N_\mathrm{conf}$ we calculate the standard deviation of the mean conductivity value:

\begin{multline}
	\Delta\sigma_\mathrm{mean}(\omega) = \frac{1}{\sqrt{N_\mathrm{conf}}}\times\\
	\sqrt{\frac{1}{N_\mathrm{conf}-1}\sum_{i=1}^{N_\mathrm{conf}}(\sigma_i(\omega)-\left\langle\sigma(\omega)\right\rangle)^2}~,
	\label{Eq:Statisticalerror}
\end{multline}
where $\sigma_i(\omega)$ is the value of the dynamic electrical conductivity for the $i^\mathrm{th}$ ionic configuration, $\left\langle\sigma(\omega)\right\rangle$ is the dynamic conductivity value averaged over configurations, $N_\mathrm{conf}$ is the number of ionic configurations. The ratio $\Delta\sigma_\mathrm{mean}(\omega) / \left\langle\sigma(\omega)\right\rangle$ is less than 2\% for all frequencies under consideration. Thus we can consider the statistical error of the conductivity calculation to be about 2\%.

We can use expression (\ref{Eq:Statisticalerror}) to estimate the statistical error only if the ionic configurations are independent. The significant decrease of the velocity-velocity autocorrelation function during the time between two subsequent configurations could be a good criterion of the independence of configurations. In this work, however, this criterion is not used.

\subsection{The dependence on the number of atoms}

\begin{figure}
	\includegraphics[width=0.95\columnwidth]{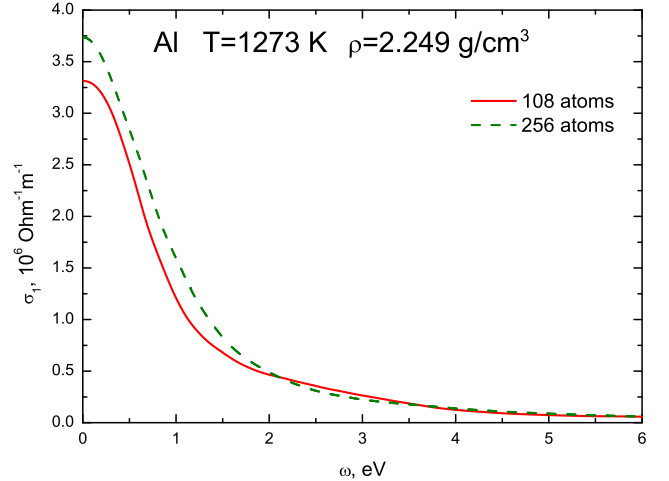}
	\caption{(Color online) The dependence of the calculated dynamic electrical conductivity on the number of atoms. Solid line --- 108 atoms; dashed line --- 256 atoms.}
	\label{Fig:Natomsdependence}
\end{figure}

The change of the number of atoms strongly influences the calculated transport and optical properties. Fig.~\ref{Fig:Natomsdependence} shows the dependence of the calculated dynamic electrical conductivity on the number of atoms in the supercell.

The change of the number of atoms from 108 to 256 leads to the significant changes in the dynamic electrical conductivity. For example for the frequency $\omega=1$~eV the conductivity changes by the factor of 1.3.  At the frequencies near to zero the change of the conductivity is about 10\%. Evidently, even larger numbers of atoms should be taken to achieve the convergence. But at the current stage it looks more reasonable to take the results obtained with the largest (256) number of atoms.

It should be mentioned, that the convergence with the number of atoms is strongly connected with the dependence of the results on the broadening of the $\delta$-function in the Kubo-Greenwood formula. This is to be discussed in the next section.

\begin{figure*}
	a)\includegraphics[width=0.95\columnwidth]{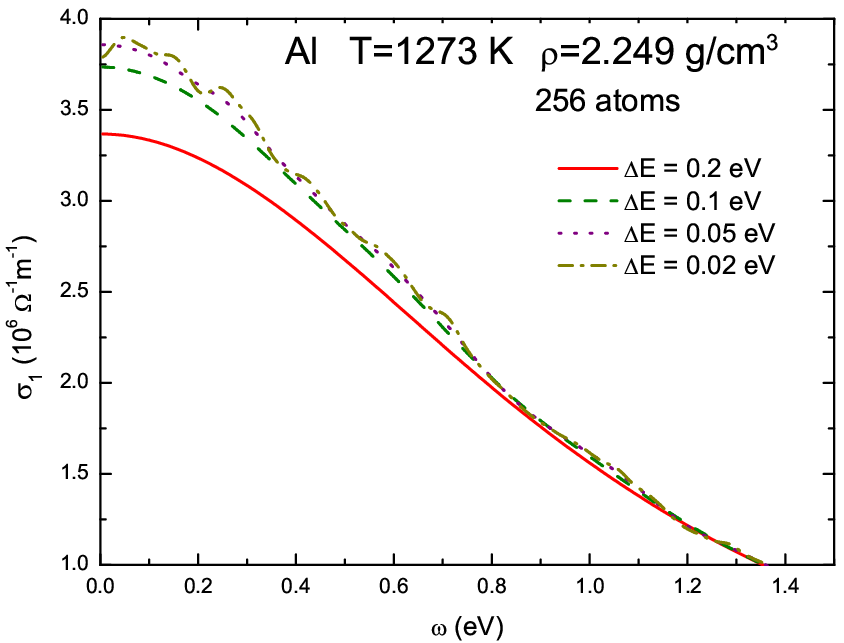}
	b)\includegraphics[width=0.95\columnwidth]{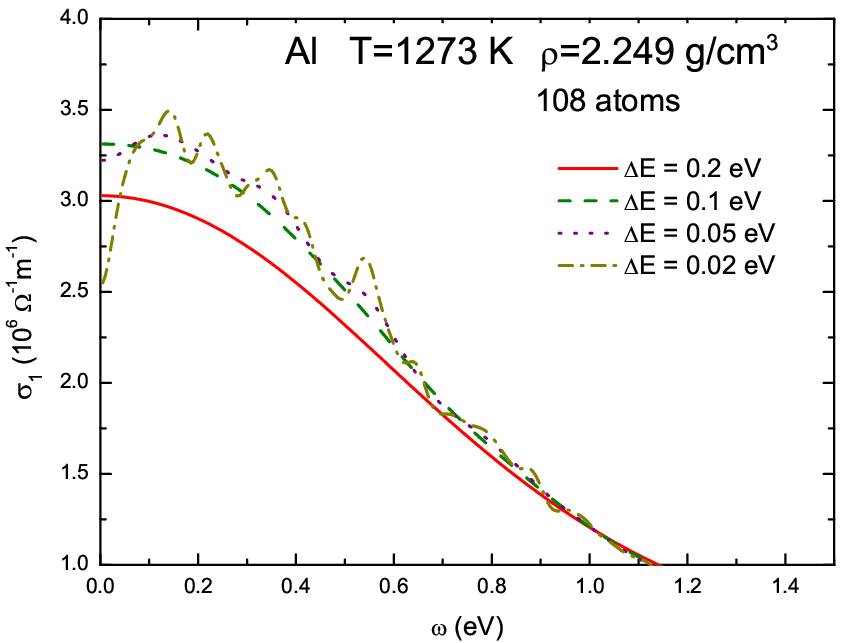}
	\caption{(Color online) The dependence of the calculated dynamic electrical conductivity on the broadening $\Delta E$ of the $\delta$-function in the Kubo-Greenwood formula. a) 256 atoms; b) 108 atoms.}
	\label{Fig:Deltaedependence}
\end{figure*}

\subsection{The dependence on the broadening of the $\delta$-function}
\label{Subsec:Technique:Broadening}

The dependence of the calculated transport and optical properties on the broadening $\Delta E$ of the $\delta$-function in the Kubo-Greenwood formula (\ref{Eq:Deltabroadening}) is rather important (especially at low frequencies). The dependence of the dynamic electrical conductivity at low frequencies on the broadening of the $\delta$-function is shown in Fig.~\ref{Fig:Deltaedependence}a.

The Kubo-Greenwood formula takes into account the transitions between all bands. In the real substance the number of atoms is very large, the bands are very close to each other. Therefore a smooth curve of dynamic electrical conductivity $\sigma_1(\omega)$ is formed. The number of atoms in numerical calculation is rather small, and the transition between each two bands gives its own $\delta$-peak to the curve of dynamic electrical conductivity. To form a smooth curve, each $\delta$-peak is to be broadened by the Gaussian function (\ref{Eq:Deltabroadening}). If the broadening $\Delta E$ is too small (0.02~eV in Fig.~\ref{Fig:Deltaedependence}a), the transitions between bands are still noticeable and the curve is oscillating. If the broadening $\Delta E$ is too large (0.2~eV in Fig.~\ref{Fig:Deltaedependence}a), too many transitions contribute to each point on the $\sigma_1(\omega)$ curve and the physical dependence on the frequency is smoothed.

We consider, that the convergence according to the broadening $\Delta E$ must occur as follows: when the broadening is diminished, at first, excessive smoothing of physical dependence disappears. Then for some $\Delta E$ value the further diminishing of the broadening does not affect the $\sigma_1(\omega)$. For some range of $\Delta E$ the curves do not depend on the broadening and any curve from this range may be taken as final. And only at very small values of $\Delta E$ the oscillations appear.

At rather high frequencies (above 1~eV in Fig.~\ref{Fig:Deltaedependence}a) the situation is similar to the described above. But at low frequencies the situation is different. When the broadening is diminished the oscillations appear earlier than the convergence of the dynamic electrical conductivity is reached.

The convergence with the broadening $\Delta E$ can be improved by the increasing of the number of atoms. When the number of atoms is relatively big, it becomes possible to take smaller value of $\Delta E$ to reach the convergence according to this parameter and avoid the appearance of oscillations. Fig.~\ref{Fig:Deltaedependence}b confirms this statement: the curves for the same values $\Delta E$ are less smooth than in Fig.~\ref{Fig:Deltaedependence}a.

At the currently available number of atoms (256, Fig.~\ref{Fig:Deltaedependence}a) the error connected with the choice of $\Delta E$ is not less than 3\% (the difference between the curves for $\Delta E=0.1$~eV and $\Delta E=0.05$~eV).

\subsection{Convergence according to the energy cut-off}
\label{Subsec:Technique:Encut}

The value of energy cut-off $E_\mathrm{cut}=200$~eV is used during the precise resolution of the band structure. The increasing of the energy cut-off to the value of 400~eV leads to the negligible (less than 1\%) change of dynamic electrical conductivity within the whole range of frequencies under consideration. On the other hand, the diminishing of the energy cut-off to the value of 100~eV leads to the excessive (up to 13\%) changes in the dynamic electrical conductivity. Therefore the value $E_\mathrm{cut}=200$~eV is optimal during the precise resolution of the band structure.

During the QMD simulation the smaller value of the energy cut-off $E_\mathrm{cut}=100$~eV is used. The increasing of the energy cut-off to the value 200~eV leads to the changes of dynamic electrical conductivity not larger than 3\% in the whole range of frequencies under consideration. On the other hand, the QMD simulation with the energy cut-off 200~eV requires significant computation time.

Therefore the error connected with energy cut-off convergence is estimated as 3\%.

\subsection{The dependence on the pseudopotential and exchange-correlation functional}
\label{Subsec:Technique:Pseudopotential}

The ultrasoft pseudopotential of Vanderbilt \cite{Vanderbilt:PRB:1990} (US-PP) in pair with local density approximation of the exchange-correlation functional (LDA) is used during QMD simulation in this work. US-PP was chosen due to its high computational efficiency.

We also use the same pair US-PP, LDA during the precise resolution of the band structure. This leads to some difficulties. At this stage we get pseudo wave functions from VASP, use them to calculate pseudo matrix elements $\left\langle\Psi_i^\mathrm{pseudo}\left|\nabla_{\alpha}\right|\Psi_j^\mathrm{pseudo}\right\rangle$, and then use them to calculate the dynamic electrical conductivity via the Kubo-Greenwood formula (\ref{Eq:KuboGreenwoodsigma}). The recalculation of pseudo matrix elements $\left\langle\Psi_i^\mathrm{pseudo}\left|\nabla_{\alpha}\right|\Psi_j^\mathrm{pseudo}\right\rangle$ to all electron matrix elements $\left\langle\Psi_i^\mathrm{AE}\left|\nabla_{\alpha}\right|\Psi_j^\mathrm{AE}\right\rangle$ is a difficult procedure for US-PP (as mentioned in \cite{Desjarlais:PRE:2002}), which is not currently implemented in the VASP package. We use parallel program module created for this work to calculate pseudo matrix elements and dynamic electrical conductivity. Currently the recalculation of pseudo matrix element to all electron is omitted.

An alternative approach is to use projected augmented-wave (PAW) pseudopotential \cite{Kresse:PRB:1999} during the precise resolution of the band structure. The VASP package provides the module that calculates all electron matrix elements $\left\langle\Psi_i^\mathrm{AE}\left|\nabla_{\alpha}\right|\Psi_j^\mathrm{AE}\right\rangle$. Unfortunately in the current VASP version this module is not parallelized. Then we use matrix elements from VASP to calculate dynamic electrical conductivity via the Kubo-Greenwood formula. In this work we perform one calculation with PAW pseudopotential in pair with Perdew, Burke, Ernzerhof (PBE) exchange-correlation functional \cite{Perdew:PRL:1996}.

The replacement of US-PP, LDA by the PAW, PBE pair gives the change of the dynamic electrical conductivity of not more than 4\% in the whole range of frequencies. The largest change (4\%) is obtained for the frequencies near to zero frequency. The small difference in results makes it possible to use US-PP in spite of difficulties described above.

\begin{figure}
	\includegraphics[width=0.95\columnwidth]{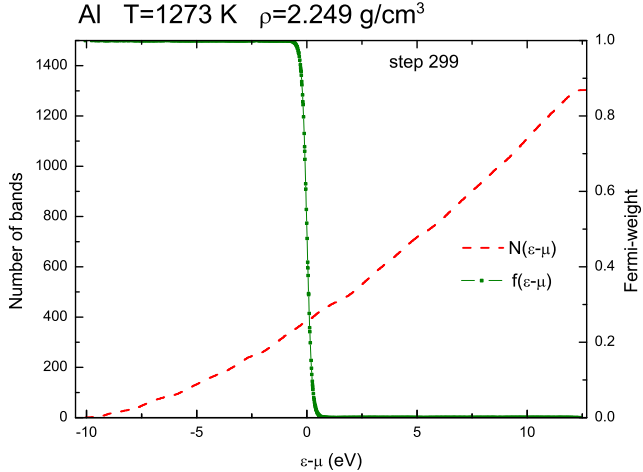}
	\caption{(Color online) The dependence of the band number (dashed line) and Fermi-weight (solid line with points) on the energy counted from the chemical potential for one ionic configuration. The points on the $f(\epsilon-\mu)$ curve show the discrete bands.}
	\label{Fig:Nbandschoice}
\end{figure}

\subsection{The choice of the number of bands}
\label{Subsec:Technique:Nbands}

During the QMD simulation all the bands with significant Fermi-weights should be taken into account. If some of the occupied bands are neglected, their contribution to electron density will be missed and ionic configurations will be calculated inaccurately. We consider the bands with the Fermi-weights larger than $10^{-5}$ (standard VASP precision during output of the Fermi-weights) to be occupied. Fig.~\ref{Fig:Nbandschoice} shows the dependence of the band number and Fermi-weight on the energy counted from the chemical potential for one ionic configuration. The Fermi-weights become less than $10^{-5}$ at $\epsilon-\mu\approx1.25$~eV. This corresponds to 500 bands.

Precise resolution of the band structure for the further calculation of the dynamic electrical conductivity requires the larger number of bands. In this work the dynamic electrical conductivity is calculated for the frequencies up to $\omega_\mathrm{max}=10$~eV. The Kubo-Greenwood formula (\ref{Eq:KuboGreenwoodsigma}) shows that the bands higher than occupied bands by the energy $\hbar\omega_\mathrm{max}=10$~eV give contribution to the value of conductivity $\sigma_1(\omega_\mathrm{max})$. Because of the broadening of the $\delta$-function this quantity should be additionally increased by $3\Delta E=0.3$~eV. For the example shown in Fig.~\ref{Fig:Nbandschoice} the bands with the energies up to $\epsilon-\mu\approx1.25~\mathrm{eV}+10~\mathrm{eV}+0.3~\mathrm{eV}\approx11.6$~eV should be taken into account. This corresponds to 1300 bands.

\subsection{Extrapolation to the zero frequency}
\label{Subsec:Technique:Extrapolation}

\begin{figure}
	\includegraphics[width=0.95\columnwidth]{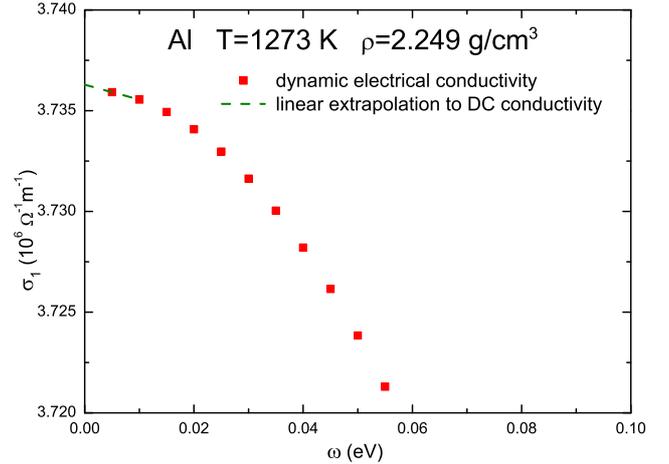}
	\caption{(Color online) The extrapolation of the dynamic electrical conductivity to the zero frequency. Square points --- calculated values of the dynamic electrical conductivity; dashed line --- linear extrapolation to the zero frequency.}
	\label{Fig:Linearextrapolation}
\end{figure}

The static value of electrical conductivity is obtained by the extrapolation of dynamic values to the zero frequency. The procedure of interpolation is shown in Fig.~\ref{Fig:Linearextrapolation}. The dynamic electrical conductivity is calculated on the dense frequency mesh (in this work the frequency step is 0.005~eV). If the number of atoms is large enough and the broadening of the $\delta$-function is chosen correctly the dynamic electrical conductivity at low frequencies is a smooth function of frequency. The linear extrapolation based on the two points nearest to the zero frequency gives the value of the static conductivity. The error of the extrapolation is negligibly small, and the error of the static value is completely determined by the error of the dynamic values used for extrapolation.

\subsection{Estimation of static electrical conductivity calculation error}
\label{Subsec:Errorestimation}

If the influence of different technical parameters on the results is examined it is possible to estimate the calculation error.

The largest contribution to the error of static electrical conductivity calculation is given by the change of the number of atoms. When the number of atoms is increased from 108 to 256 the static electrical conductivity grows by 10\%. The behavior of the static electrical conductivity under the further increasing of the number of atoms is not clear. But taking into account available data the error connected with the number of atoms may be estimated as +10\%.

The replacement of the US-PP, LDA pair by the PAW, PBE pair leads to the decrease of the static electrical conductivity by 4\%. The error connected with the pseudopotential and exchange-correlation functional choice can be estimated as $\pm4\%$.

The error connected with the energy cut-off choice is about $\pm3\%$.

The choice of the broadening of the $\delta$-function $\Delta E$ yields the error +3\%.

The statistical error is $\pm2\%$.

The error connected with the extrapolation of the dynamic electrical conductivity to zero frequency is negligibly small.

Unfortunately, currently the dependence of the results on the number of \textbf{k}-points is not examined because of too large computational requirements (see, however, the analysis of computational errors in \ref{Sec:Appendix}).

We estimate total error by the simple algebraic sum of errors from different sources. Thus we estimate the range where the true value of static electrical conductivity may be. The factors increasing the value of static electrical conductivity give the total error of $10\%+4\%+3\%+3\%+2\%=22\%$. The factors decreasing the value of static electrical conductivity give the total error of $4\%+3\%+2\%=9\%$.

We consider that the precision of the calculation of static electrical conductivity can be improved by further increasing of the number of atoms and by the examination of the convergence with the number of \textbf{k}-points.

\begin{figure}
	\includegraphics[width=0.95\columnwidth]{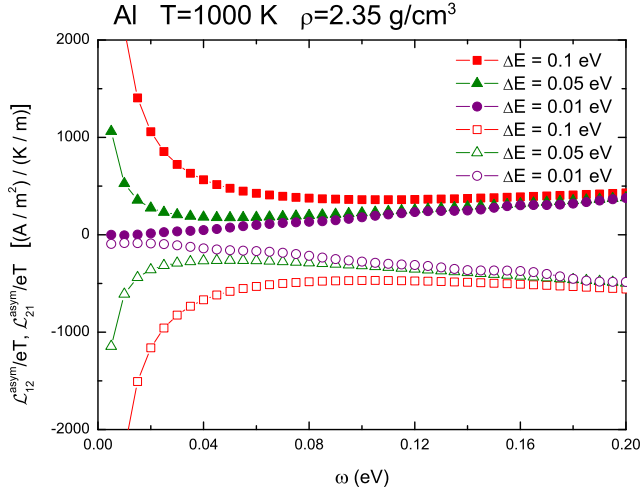}
	\caption{(Color online) The dynamic Onsager coefficients $\mathcal{L}_{12}^\mathrm{asym}(\omega)/eT$ and $\mathcal{L}_{21}^\mathrm{asym}(\omega)/eT$ calculated according to the asymmetric formula (\ref{Eq:DynamicOnsagerAsym}) for the different broadenings of the $\delta$-function. Filled symbols~--- $\mathcal{L}_{12}^\mathrm{asym}(\omega)/eT$, empty symbols~--- $\mathcal{L}_{21}^\mathrm{asym}(\omega)/eT$.}
	\label{Fig:OnsagerL12L21Asym}
\end{figure}

\section{DYNAMIC ONSAGER COEFFICIENTS}
\label{Sec:Onsager}

In this section we discuss different expressions for calculation of the dynamic Onsager coefficients. The comparison is performed for liquid aluminum at $T=1000$~K and $\rho=2.35$~g/cm$^3$. The technical parameters are the same as described in section \ref{Sec:Technique}, except for the broadening of the $\delta$-function $\Delta E=0.07$~eV.

The Kubo-Greenwood formula for the calculation of the dynamic Onsager coefficients (\ref{Eq:DynamicOnsager}) is derived in the paper \cite{Holst:PRB:2011}. The $\mathcal{L}_{11}(\omega)$ Onsager coefficient is the dynamic electrical conductivity $\sigma_1(\omega)$. Other dynamic Onsager coefficients in this work have no direct physical meaning and are used only to obtain static values by the extrapolation to the zero frequency. In the limit of the zero frequency the Kubo-Greenwood formula (\ref{Eq:DynamicOnsager}) reduces to:
\begin{multline}
\mathcal{L}_{mn}=(-1)^{m+n}\frac{2\pi e^2\hbar^3}{3m^2\Omega}\times\\
\sum_{i,j,\alpha,\mathbf{k}}W(\mathbf{k})(\epsilon_{i,\mathbf{k}}-\mu)^{m+n-2}\left|\left\langle\Psi_{i,\mathbf{k}}\left|\nabla_\alpha\right|\Psi_{j,\mathbf{k}}\right\rangle\right|^2\times\\
\left(-\frac{\partial f}{\partial\epsilon}\right)_{\epsilon=\epsilon_{i,\mathbf{k}}}\delta(\epsilon_{j,\mathbf{k}}-\epsilon_{i,\mathbf{k}}).
\label{Eq:OnsagerStatic}
\end{multline}
 
Here the expression $\frac{f(\epsilon_{i,\mathbf{k}})-f(\epsilon_{j,\mathbf{k}})}{\hbar\omega}$ in the Kubo-Greenwood formula (\ref{Eq:DynamicOnsager}) tends to $\left(-\frac{\partial f}{\partial\epsilon}\right)_{\epsilon=\epsilon_{i,\mathbf{k}}}$ in the limit of zero frequency. Also due to the $\delta$-function in the formula (\ref{Eq:OnsagerStatic}) the values $\epsilon_{i,\mathbf{k}}$ and $\epsilon_{j,\mathbf{k}}$ are equal and there is no difference between using the factor of $(\epsilon_{i,\mathbf{k}}-\mu)^{m+n-2}$ or $(\epsilon_{j,\mathbf{k}}-\mu)^{m+n-2}$.

If the dynamic Onsager coefficients are treated only as some expressions which give correct values in the limit of zero frequency, different expressions for the dynamic Onsager coefficients may be chosen. In paper \cite{Recoules:PRB:2005} the asymmetric expression is used:
\begin{multline}
\mathcal{L}_{mn}^\mathrm{asym}(\omega)=(-1)^{m+n}\frac{2\pi e^2\hbar^2}{3m^2\omega\Omega}\times\\
\sum_{i,j,\alpha,\mathbf{k}}W(\mathbf{k})(\epsilon_{i,\mathbf{k}}-\mu)^{n-1}(\epsilon_{j,\mathbf{k}}-\mu)^{m-1}
\left|\left\langle\Psi_{i,\mathbf{k}}\left|\nabla_\alpha\right|\Psi_{j,\mathbf{k}}\right\rangle\right|^2\times\\
\left[f(\epsilon_{i,\mathbf{k}})-f(\epsilon_{j,\mathbf{k}})\right]\delta(\epsilon_{j,\mathbf{k}}-\epsilon_{i,\mathbf{k}}-\hbar\omega)
\label{Eq:DynamicOnsagerAsym}
\end{multline} 
that gives correct zero frequency limit (\ref{Eq:OnsagerStatic}).
 
The Onsager coefficients $\mathcal{L}_{12}^\mathrm{asym}(\omega)/eT$ and $\mathcal{L}_{21}^\mathrm{asym}(\omega)/eT$ calculated according to the formula (\ref{Eq:DynamicOnsagerAsym}) are shown in Fig.~\ref{Fig:OnsagerL12L21Asym}. The coefficients are not equal at non-zero frequencies. At the reasonable (see~\ref{Subsec:Technique:Broadening}) broadenings of the $\delta$-function 0.1~eV and 0.05~eV $\mathcal{L}_{12}^\mathrm{asym}(\omega)/eT$ and $\mathcal{L}_{21}^\mathrm{asym}(\omega)/eT$ coefficients diverge at the zero frequency limit. And only at very small broadening 0.01~eV two curves come to the close values at the zero frequency limit. Such a small value of broadening can not be used because of too large oscillations on the $\sigma_1(\omega)$ (see~\ref{Subsec:Technique:Broadening}).

\begin{figure}
	\includegraphics[width=0.95\columnwidth]{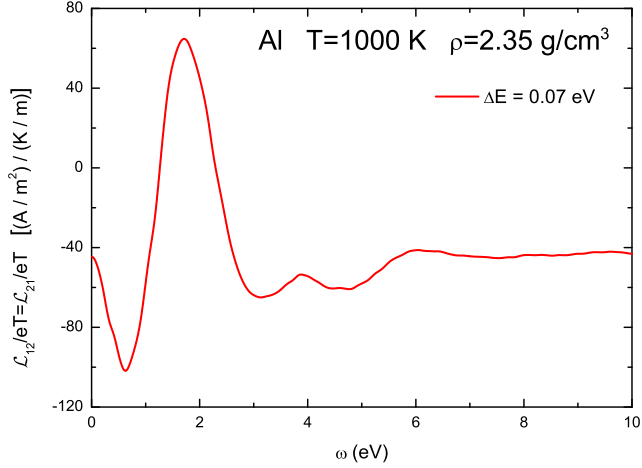}
	\caption{(Color online) The frequency dependence of the dynamic Onsager coefficients $\mathcal{L}_{12}(\omega)/eT=\mathcal{L}_{21}(\omega)/eT$ calculated according to the symmetric expression (\ref{Eq:DynamicOnsager}).}
	\label{Fig:OnsagerL12Sym}
\end{figure}

It can be easily seen from Fig.~\ref{Fig:OnsagerL12L21Asym} that the curves $\mathcal{L}_{12}^\mathrm{asym}(\omega)/eT$ and $\mathcal{L}_{21}^\mathrm{asym}(\omega)/eT$ at the zero frequency limit have approximately the same values and opposite signs. If added and divided by two they compensate each other and form the smooth curve at the zero frequency limit. The half-sum $\frac{\mathcal{L}_{12}^\mathrm{asym}(\omega)+\mathcal{L}_{21}^\mathrm{asym}(\omega)}{2}$ gives the same value as the symmetric expression (\ref{Eq:DynamicOnsager}). The values of the Onsager coefficients $\mathcal{L}_{12}(\omega)/eT=\mathcal{L}_{21}(\omega)/eT$ calculated according to the symmetric expression (\ref{Eq:DynamicOnsager}) for $\Delta E=0.07$~eV are shown in Fig.~\ref{Fig:OnsagerL12Sym}. In this case the curve is smooth at the zero frequency limit.

\begin{figure}
	\includegraphics[width=0.95\columnwidth]{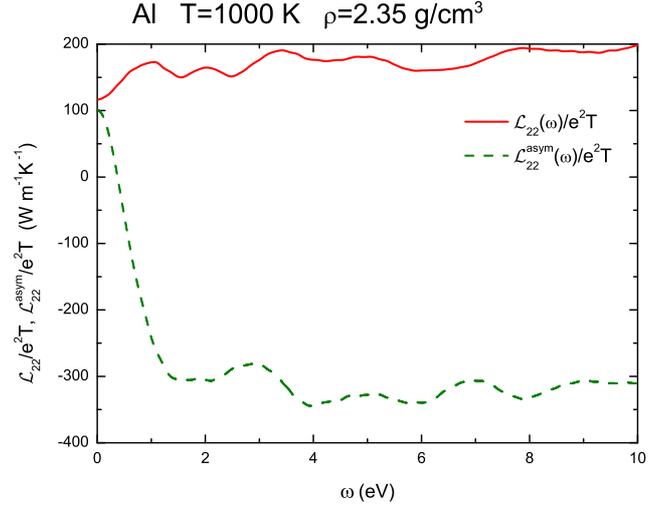}
	\caption{(Color online) The frequency dependence of the Onsager coefficient $\mathcal{L}_{22}(\omega)/e^2T$ calculated according to the symmetric (solid line, equation (\ref{Eq:DynamicOnsager})) and asymmetric (dashed line, equation (\ref{Eq:DynamicOnsagerAsym})) expressions. The broadening of the $\delta$-function is 0.07~eV.}
	\label{Fig:OnsagerL22SymAsym}
\end{figure}

As concerns the $\mathcal{L}_{22}(\omega)$ Onsager coefficient, its frequency dependences calculated according to the symmetric (\ref{Eq:DynamicOnsager}) and asymmetric (\ref{Eq:DynamicOnsagerAsym}) expression are shown in Fig.~\ref{Fig:OnsagerL22SymAsym}. In the case of the $\mathcal{L}_{22}(\omega)$ coefficient, the asymmetric curve is smooth at the limit of zero frequency. The symmetric and asymmetric expressions give the close static values, although the symmetric value is 20\% higher than asymmetric one.

Thus, both symmetric (\ref{Eq:DynamicOnsager}) and asymmetric (\ref{Eq:DynamicOnsagerAsym}) expressions for the dynamic Onsager coefficients theoretically give the correct zero frequency limit (\ref{Eq:OnsagerStatic}). But the symmetric expression is more numerically stable and gives no divergence at zero frequency limit at the numerical calculation of $\mathcal{L}_{12}(\omega)$ and $\mathcal{L}_{21}(\omega)$. Also it is more justified theoretically \cite{Holst:PRB:2011}. In this work we use only the symmetric expression (\ref{Eq:DynamicOnsager}) to calculate the dynamic Onsager coefficients.

\section{RESULTS AND DISCUSSION}
\label{Sec:Results}

\subsection{Dynamic electrical conductivity and optical properties}

\subsubsection{The comparison with other calculations}

\begin{figure}
	a)\includegraphics[width=0.95\columnwidth]{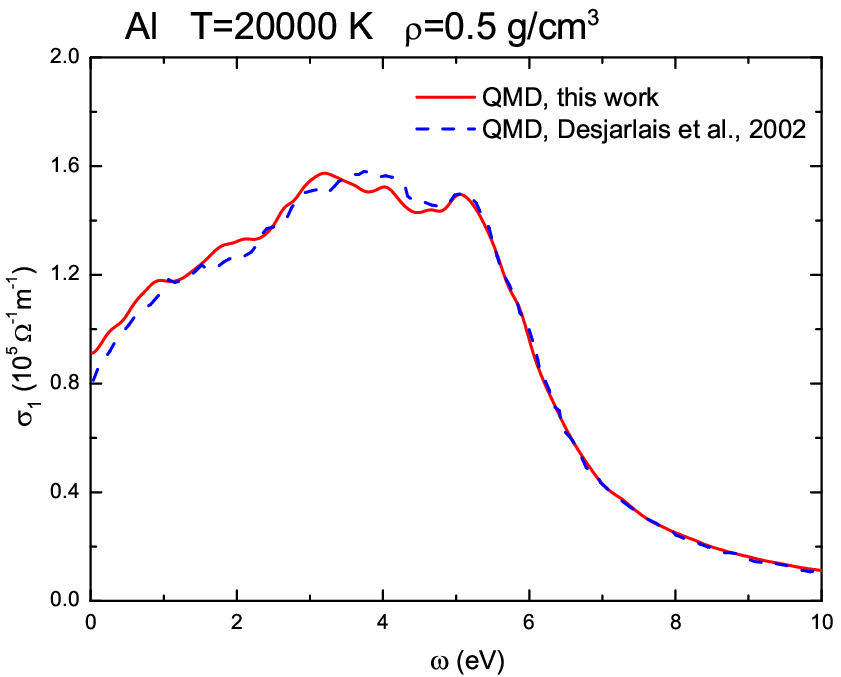}
	b)\includegraphics[width=0.95\columnwidth]{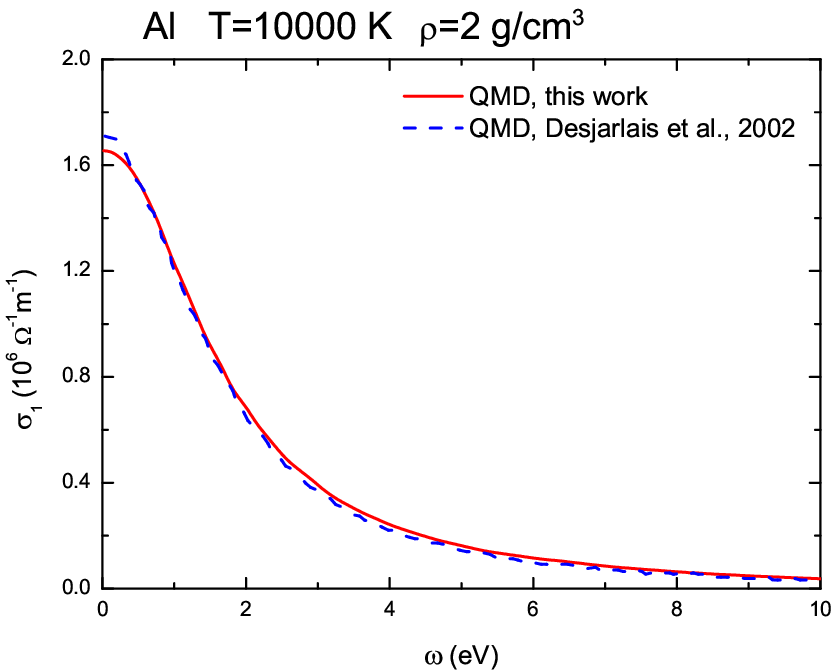}
	\caption{(Color online) The dynamic electrical conductivity of aluminum in comparison with calculation \cite{Desjarlais:PRE:2002}. Solid line~--- this work, dashed line~--- calculation \cite{Desjarlais:PRE:2002}. The calculation parameters: a) 32 atoms; PAW pseudopotential; PBE exchange-correlation functional; 1 \textbf{k}-point in the Brillouin zone; energy cut-off 200~eV; the broadening of the $\delta$-function 0.1~eV; 15 ionic configurations; 1500 steps; 1 step -- 2~fs. b) 108 atoms; US-PP pseudopotential; LDA exchange-correlation functional; 1 \textbf{k}-point in the Brillouin zone; energy cut-off 100~eV; the broadening of the $\delta$-function 0.1~eV; 15 ionic configurations; 1500 steps; 1 step -- 2~fs.}
	\label{Fig:Desjarlais}
\end{figure}

We perform the calculation at the same temperatures and densities as in the well-known paper of Desjarlais \cite{Desjarlais:PRE:2002} to test our computational scheme. The technical parameters of calculation are taken close to those of Desjarlais. The results are presented in Fig.~\ref{Fig:Desjarlais}. The good agreement means that the calculation performed previously may be well reproduced using the same computational package and the close values of the parameters.

\subsubsection{Normal isochor}

\begin{figure}
	\includegraphics[width=0.95\columnwidth]{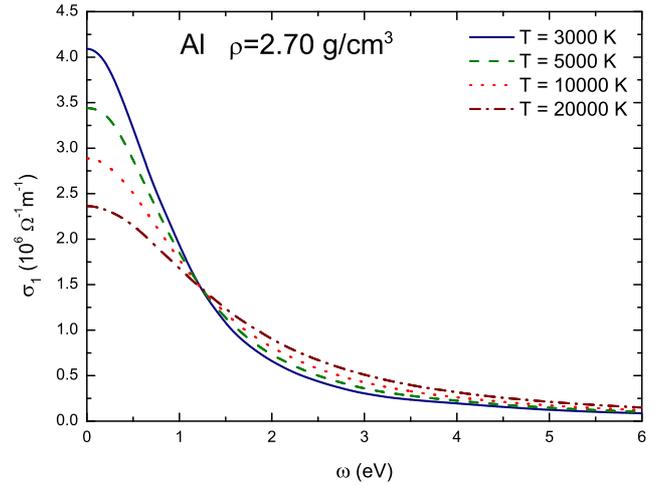}
	\caption{(Color online) The frequency dependences of the dynamic electrical conductivity at the normal isochor for different temperatures. The calculation parameters are the same as described in section~\ref{Sec:Technique}.}
	\label{Fig:Normalisochor}
\end{figure}

The results on the dynamic electrical conductivity at the normal-density isochor are shown in Fig.~\ref{Fig:Normalisochor}. The frequency dependence of the dynamic electrical conductivity has the Drude-like shape. The dynamic electrical conductivity decreases with the temperature growth at the low frequencies, and, on the contrary increases at high frequencies. Also, it is interesting to mention, that the dynamic electrical conductivity does not depend on the temperature at $\omega\approx1.25$~eV.

\subsubsection{Dynamic Onsager coefficients}

In this work only the frequency dependence of the $\mathcal{L}_{11}(\omega)$ Onsager coefficient has the direct physical meaning of the real part of dynamic electrical conductivity $\sigma_1(\omega)$. The frequency dependences of the other Onsager coefficients are used only to obtain the static values by the extrapolation to the zero frequency. Nevertheless, we consider that it may be useful to show also the curves of the dynamic Onsager coefficients to understand the procedure of the extrapolation to the zero frequency.

\begin{figure}
	a)\includegraphics[width=0.95\columnwidth]{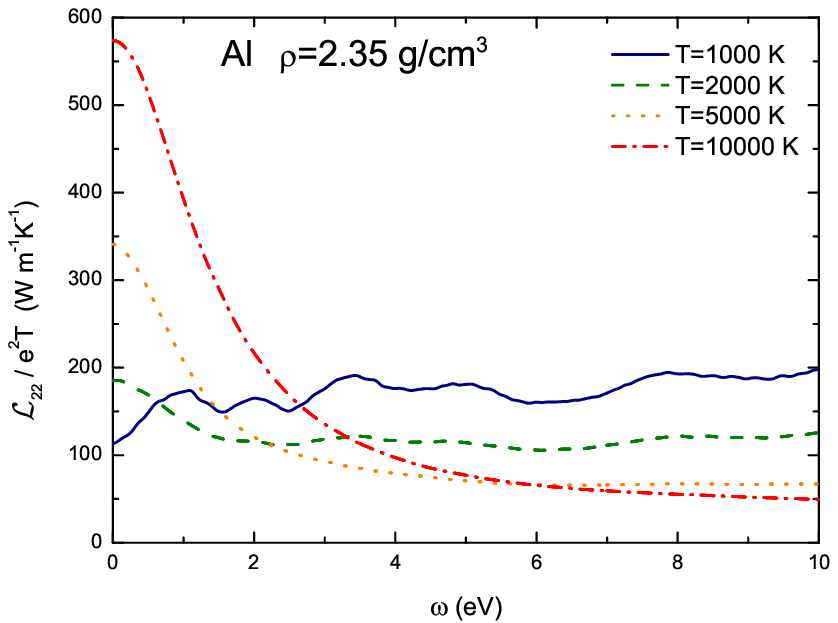}
	b)\includegraphics[width=0.95\columnwidth]{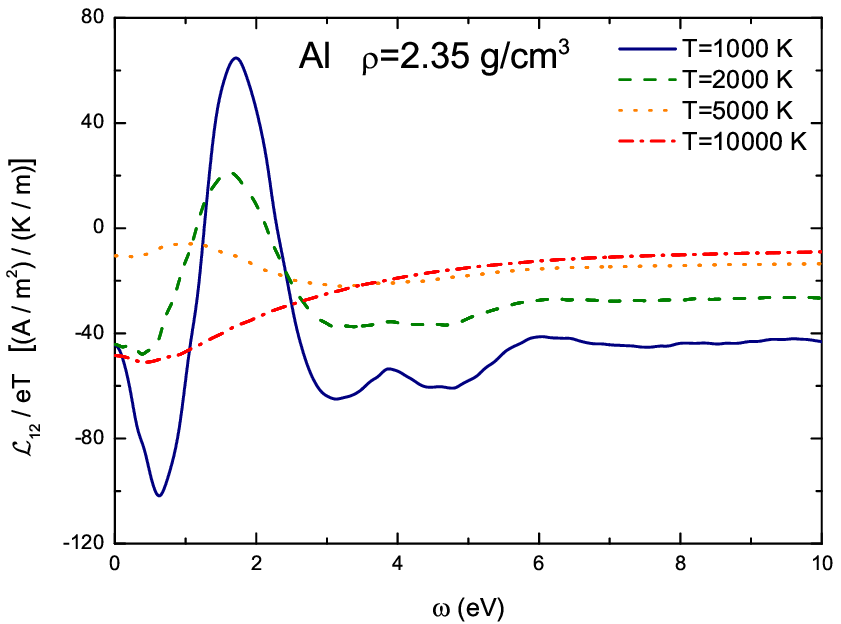}
	\caption{(Color online) The frequency dependences of the Onsager coefficients $\mathcal{L}_{22}(\omega)/e^2T$ (a) and $\mathcal{L}_{12}(\omega)/eT$ (b) for aluminum, $\rho=2.35$~g/cm$^3$. The technical parameters of the calculation are the same as in section~\ref{Sec:Technique} (except for the broadening of the $\delta$-function at temperatures up to 2000~K, $\Delta E=0.07$~eV).}
	\label{Fig:DynamicOnsager}
\end{figure}

Fig.~\ref{Fig:DynamicOnsager}a shows the frequency dependences of the Onsager coefficient $\mathcal{L}_{22}(\omega)/e^2T$ at density $\rho=2.35$~g/cm$^3$ for different temperatures from 1000~K up to 10000~K. The static value $\mathcal{L}_{22}(0)/e^2T$ is the approximate value of the thermal conductivity (neglecting the thermoelectric term, (\ref{Eq:KApprox})). The shape of the curve $\mathcal{L}_{22}(\omega)/e^2T$ changes significantly with the temperature increasing. Nevertheless the static values form the smooth dependence on the temperature (see~\ref{Subsubsec:Results:Static:Isochor2.35}).

Fig.~\ref{Fig:DynamicOnsager}b presents the frequency dependences of the Onsager coefficient $\mathcal{L}_{12}(\omega)/eT$ at the isochor $\rho=2.35$~g/cm$^3$ for the temperatures from 1000~K up to 10000~K. The static value $\mathcal{L}_{12}(0)/eT$ is necessary to calculate thermoelectric term in the expression (\ref{Eq:KExact}). The shape of the curve changes significantly with the temperature growth. The static values form the nonmonotonic dependence on the temperature. Also for all the temperatures less than 10000~K the thermoelectric term is negligibly small (see~\ref{Subsubsec:Results:Static:Normisochor}). These two facts mean that additional investigation should be performed to get precise values $\mathcal{L}_{12}(0)/eT$.

\subsection{Static electrical and thermal conductivity}

\subsubsection{Normal isobar}

\begin{figure}
	\includegraphics[width=0.95\columnwidth]{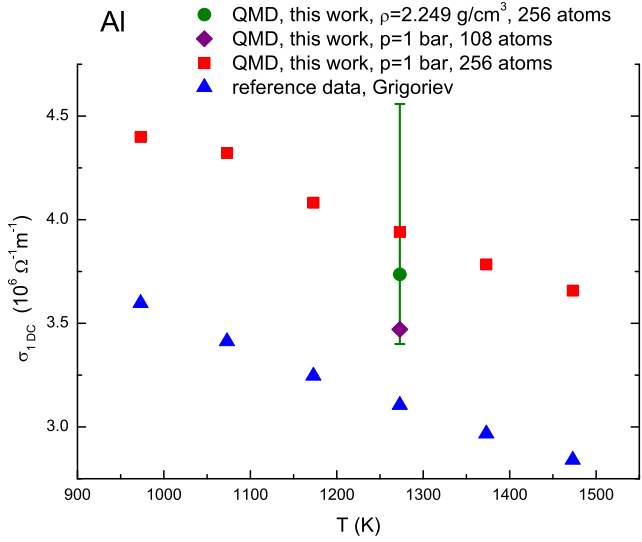}
	\caption{(Color online) The comparison of the calculated static electrical conductivity on the normal isobar with the reference data \cite{Grigoriev:Englishlink:1991}. Circle with error bars --- the calculated point $T=1273$~K, $\rho=2.249$~g/cm$^3$, the error estimation for this point is performed in section~\ref{Sec:Technique}, 256 atoms. Diamond  --- the calculated point at the normal isobar with the decreased number of atoms --- 108. Squares --- the calculated points at the normal isobar, 256 atoms. Triangles --- reference data \cite{Grigoriev:Englishlink:1991}. The technical parameters are the same as described in section~\ref{Sec:Technique} (except for the point with 108 atoms).}
	\label{Fig:Normalisobar}
\end{figure}

The calculation for liquid aluminum at the normal isobar for the temperatures from 973~K up to 1473~K was made to perform comparison with the reference data. The values of the density of liquid aluminum at the normal isobar, necessary during the QMD simulation are taken from \cite{Lide:2005}. The results of the calculation in comparison with reference data are shown in Fig.~\ref{Fig:Normalisobar}.

The reference density of liquid aluminum at normal pressure and temperature $T=1273$~K is $\rho=2.296$~g/cm$^3$. The error estimation in section~\ref{Sec:Technique} is performed for temperature $T=1273$~K and density $\rho=2.249$~g/cm$^3$, that only slightly differs from the density at the normal isobar. Therefore, this error estimation is applicable for the results at the normal isobar, too.

The discrepancy between the calculated values and reference data is $\sim25\%$. The factors which increase the static electrical conductivity yield total error $+22\%$. The factors which decrease the static electrical conductivity yield total error $-9\%$. Therefore the reason of the discrepancy is still not clarified. Nevertheless, negative slopes of the static electrical conductivity curves are close for the calculated and reference data (see Fig.\ref{Fig:Normalisobar}).
 
The static electrical conductivity undergoes the most significant changes under the varying of the number of atoms (see~\ref{Subsec:Errorestimation}). The result for less number of atoms --- 108 is also shown in Fig.~\ref{Fig:Normalisobar}, the difference from the reference data is $\sim12\%$. Nevertheless we think it is more correct to consider the result with the larger number of atoms as preliminary in spite of its larger discrepancy with reference data. We consider that the further increasing of the number of atoms and the investigation of the convergence according to the number of \textbf{k}-points may improve the results.

\begin{figure}
	a)\includegraphics[width=0.95\columnwidth]{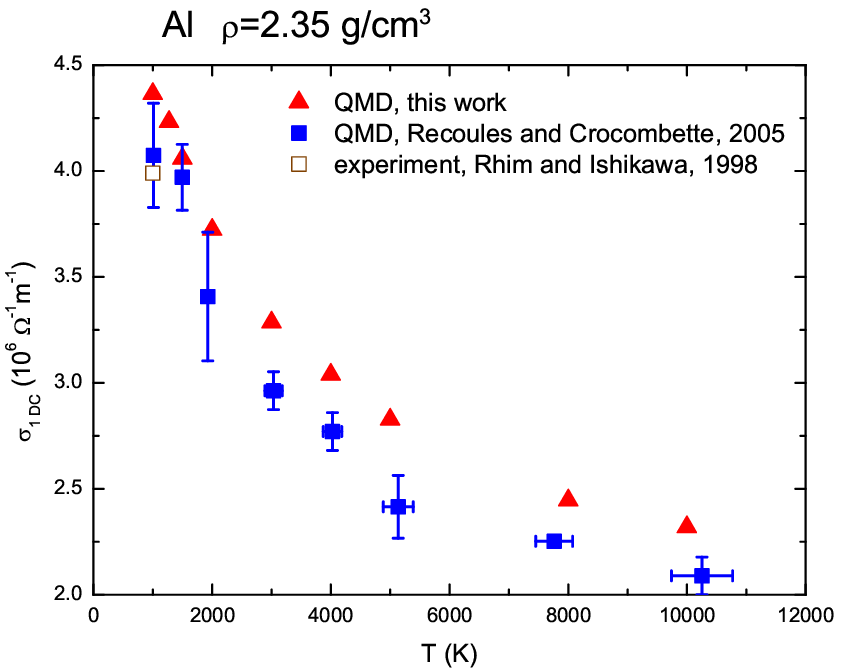}
	b)\includegraphics[width=0.95\columnwidth]{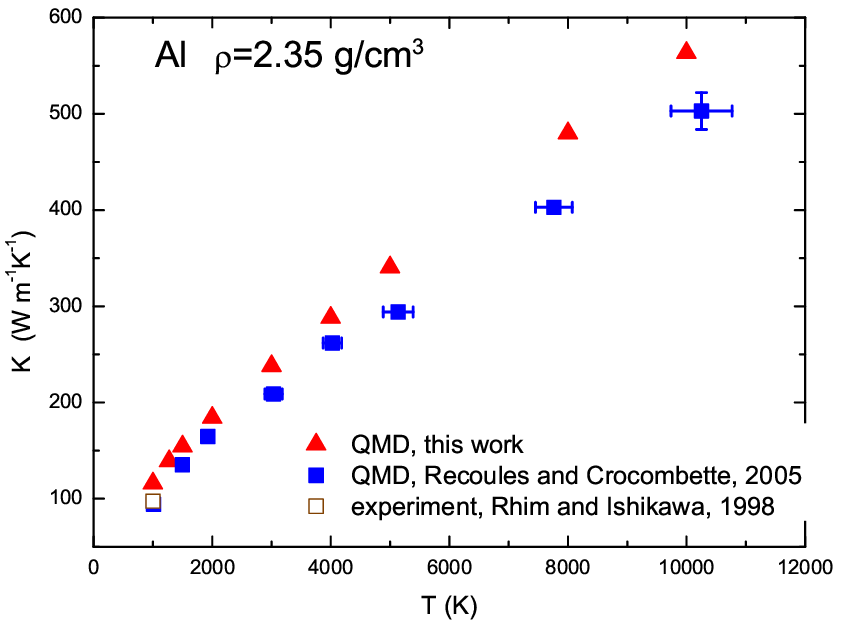}
	\caption{(Color online) The calculated static electrical conductivity (a) and thermal conductivity (b) in comparison with the results of other authors and experimental data along the isochor $\rho=2.35$~g/cm$^3$. Filled triangles --- the calculation of this work; filled squares --- calculation \cite{Recoules:PRB:2005}; empty square --- experiment \cite{Rhim:RSI:1998}. The calculation parameters are the same as in section~\ref{Sec:Technique} (except for the broadening of the $\delta$-function at temperatures up to 2000~K, $\Delta E=0.07$~eV).}
	\label{Fig:RecoulesResigmaK}
\end{figure}

\subsubsection{The isochor  2.35~g/cm$^\textit{3}$}
\label{Subsubsec:Results:Static:Isochor2.35}

To compare our results with the results of other authors the static electrical conductivity and thermal conductivity of liquid aluminum at density $\rho=2.35$~g/cm$^3$ were calculated. The temperature was varied from 1000~K up to 10000~K. The results of the calculation in comparison with the results of other authors \cite{Recoules:PRB:2005} and experimental data \cite{Rhim:RSI:1998} are shown in Fig.~\ref{Fig:RecoulesResigmaK}.

The static electrical conductivity decreases with the temperature growth, whereas thermal conductivity increases. For all the temperatures up to 10000~K the thermoelectric term in the equation (\ref{Eq:KExact}) gives small (not larger than 2\%) contribution to the thermal conductivity and the approximate formula (\ref{Eq:KApprox}) is valid. The calculated values of the static electrical conductivity are larger than in the calculation \cite{Recoules:PRB:2005} and experiment \cite{Rhim:RSI:1998}. We explain this in the same way as for the normal isobar: we use the larger number of atoms (256 atoms) than in the work \cite{Recoules:PRB:2005} (108 atoms). So the static electrical conductivity grows away from the calculation with the smaller number of atoms \cite{Recoules:PRB:2005} and experiment. At relatively low temperatures (up to 2000~K) the calculation conditions are close to that at the normal isobar, so the error estimation of the section~\ref{Sec:Technique} is applicable here.

As far as the thermal conductivity is concerned its values are also higher than in the calculation \cite{Recoules:PRB:2005} and experiment \cite{Rhim:RSI:1998}. The dependence of the thermal conductivity on the technical parameters is not investigated in this work and we can only suppose that the total error of the thermal conductivity calculation is close to that of the static electrical conductivity.

\begin{figure}
	\includegraphics[width=0.95\columnwidth]{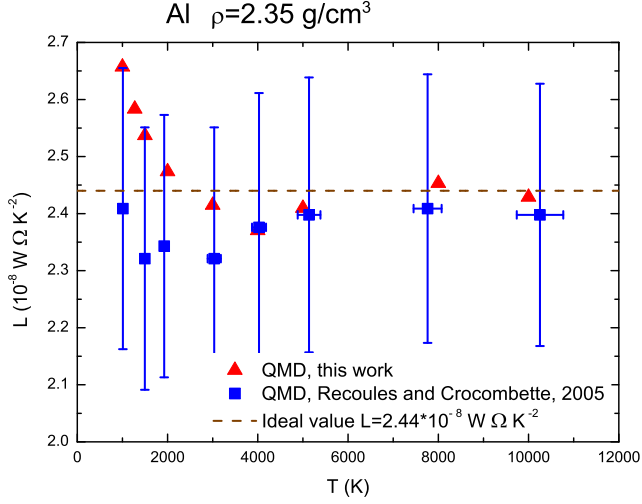}
	\caption{(Color online) The Lorenz number at the isochor $\rho=2.35$~g/cm$^3$. Triangles --- calculation of this work; squares with error bars --- calculation \cite{Recoules:PRB:2005}; dashed line --- the ideal value $L=2.44\cdot10^{-8}$~W$\cdot\Omega\cdot$K$^{-2}$.}
	\label{Fig:RecoulesLorenz}
\end{figure}

The calculated values of the Lorenz number in comparison with the calculation \cite{Recoules:PRB:2005} and ideal value $L=2.44\cdot10^{-8}$~W$\cdot\Omega\cdot$K$^{-2}$ are shown in Fig.~\ref{Fig:RecoulesLorenz}. The difference between the calculated Lorenz number, results of \cite{Recoules:PRB:2005} and ideal value is not larger than the estimated calculation error $\sim20\%$. The value of the Lorenz number for the experimental data \cite{Rhim:RSI:1998} is ideal because in that work the thermal conductivity was recalculated from the electrical conductivity via the Wiedemann-Franz law.

\subsubsection{Normal isochor}
\label{Subsubsec:Results:Static:Normisochor}

\begin{figure}
	a)\includegraphics[width=0.95\columnwidth]{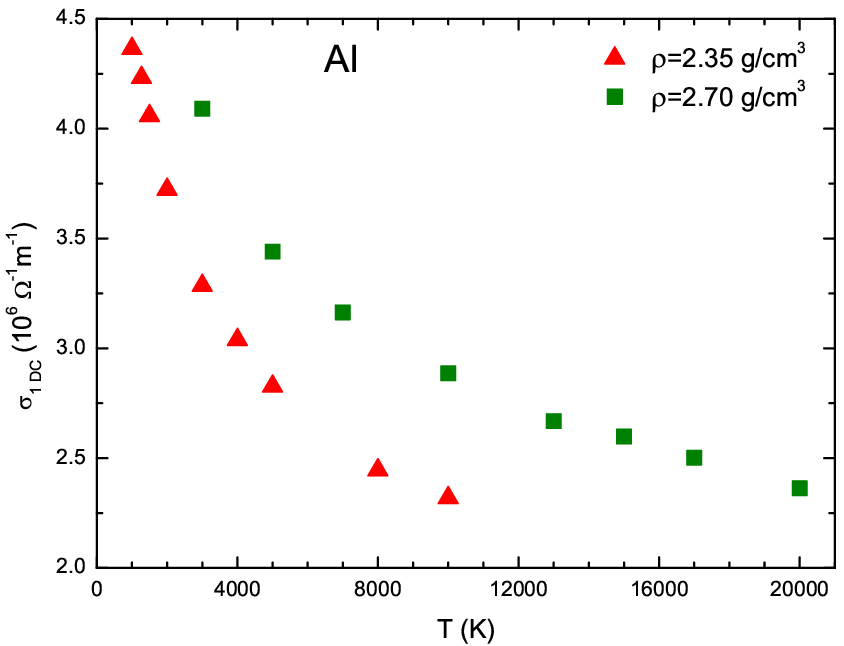}
	b)\includegraphics[width=0.95\columnwidth]{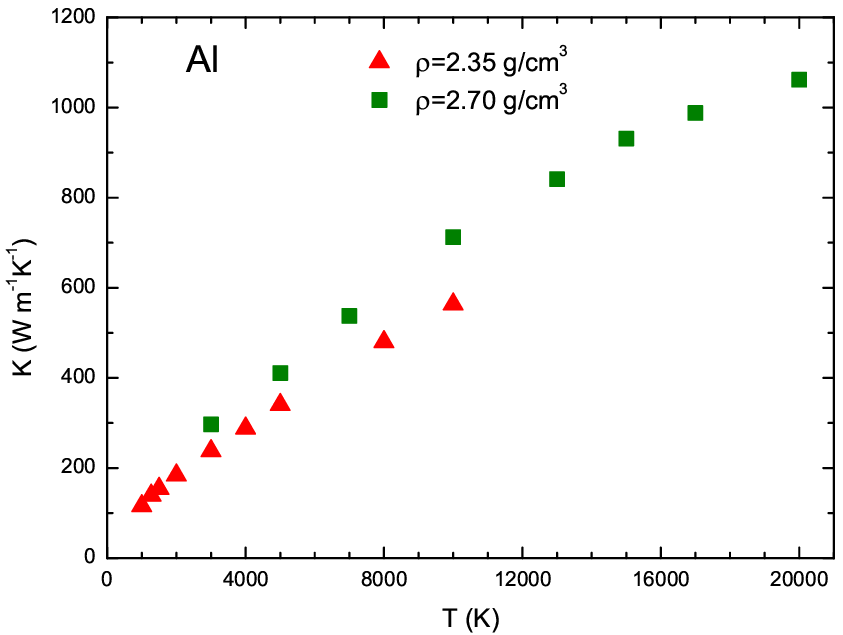}
	\caption{(Color online) The calculated static electrical conductivity (a) and thermal conductivity (b) of aluminum. The technical parameters of the calculation are the same as described in section~\ref{Sec:Technique} (except for the broadening of the $\delta$-function at the temperatures up to 2000~K, $\Delta E=0.07$~eV).}
	\label{Fig:NormalisochorDC}
\end{figure}

The results of the calculation at the normal isochor $\rho=2.70$~g/cm$^3$ and the isochor $\rho=2.35$~g/cm$^3$ are shown in Fig.~\ref{Fig:NormalisochorDC}.

Qualitatively the dependences of static electrical conductivity and thermal conductivity on the temperature at the density $\rho=2.70$~g/cm$^3$ have the same shape as at the density $\rho=2.35$~g/cm$^3$. The static electrical conductivity and thermal conductivity decrease with the growth of density.

\begin{figure}
	\includegraphics[width=0.95\columnwidth]{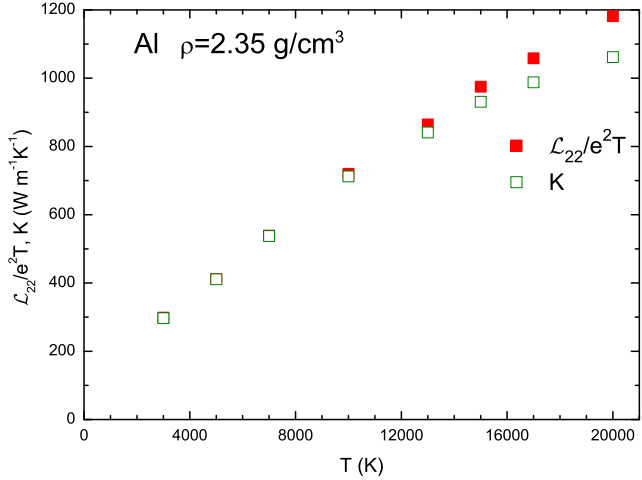}
	\caption{(Color online) The influence of the thermoelectric term on the thermal conductivity. Filled squares --- $\mathcal{L}_{22}/e^2T$, thermal conductivity calculated according to the approximate expression (\ref{Eq:KApprox}). Empty squares --- thermal conductivity calculated according to the exact expression (\ref{Eq:KExact}).}
	\label{Fig:L22K}
\end{figure}

The dependences of the thermal conductivity on the temperature calculated according to the approximate expression (\ref{Eq:KApprox}) and the exact expression (\ref{Eq:KExact}) taking into account the thermoelectric term are shown in Fig.~\ref{Fig:L22K}. The relative contribution of the thermoelectric term increases with the temperature growth. At the temperatures less than 1000 K its contribution is less than 1\% and approximate expression (\ref{Eq:KApprox}) is valid. At the maximum temperature under consideration $T=20000$~K the relative contribution of the thermoelectric term is 10\%.

\section{CONCLUSIONS}

1. The influence of the main technical parameters on the results was investigated for liquid aluminum at $T=1273$~K and $\rho=2.249$~g/cm$^3$. It was shown, that the current number of atoms --- 256 is still too small for the convergence according to the number of atoms to be achieved. The error of the calculation of the static electrical conductivity was estimated to be about 22\%. The discrepancy with the reference data is about 25\%. Further increasing of the number of atoms and the investigation of the convergence with the number of \textbf{k}-points is necessary.

2. The expressions for the dynamic Onsager coefficients from the articles \cite{Holst:PRB:2011} and \cite{Recoules:PRB:2005} were compared. The symmetric expression from the article \cite{Holst:PRB:2011} being theoretically justified was found out to be more numerically stable.

3. The calculations of the dynamic and static electrical conductivity as well as thermal conductivity were performed for liquid aluminum at near-normal densities for the temperatures form melting up to 20000~K. The results are in satisfactory agreement with reference and experimental data and the calculations of other authors.

\section*{ACKNOWLEDGEMENTS}

This work was supported by the FAIR-Russia Research Center Grant for master students (2011-2012), FAIR-Russia Research Center Grant for PhD students (2012-2013) and the Russian Foundation for Basic Research, grants 13-02-91057-CNRS and 13-08-01179.





\appendix
\section{Influence of the number of \textbf{k}-points}\label{Sec:Appendix}

During the preparation of this article we obtained some additional information concerning the dependence of the results on the number of \textbf{k}-points in the Brillouin zone during precise resolution of band structure. The calculation was performed for liquid aluminum for the same $(\rho, T)$-point as in section \ref{Sec:Technique}: $\rho=2.249$~g/cm$^3$ and $T=1273$~K. Due to huge computational time it is difficult to conduct this investigation for the supercell with 256 atoms. Nevertheless, computations were performed for 108 atoms in the supercell. All other technical parameters are the same as described in section \ref{Sec:Technique}.

In this Appendix firstly the meshes in the Brillouin zone are described. Then the procedure of reduction of the number of \textbf{k}-points due to symmetry reasons is introduced which allows to shorten significantly computational time with nearly negligible influence on the results. Further the obtained results are discussed. Though the convergence with the number of \textbf{k}-points was not achieved, the effects of the increment of the number of \textbf{k}-points may be observed. The changes in the static electrical conductivity caused by the introduction of denser meshes in the Brillouin zone are estimated.

\subsection{\textbf{k}-point meshes}

Two types of \textbf{k}-point meshes based on the Monkhorst-Pack scheme \cite{Monkhorst:PRB:1976} were used.

The first scheme is the $\Gamma$-centered Monkhorst-Pack scheme. The coordinates of \textbf{k}-points in the Brillouin zone for the cubic supercell with the side $a$ are given by:

\begin{equation}
k_x=\frac{2\pi}{a}\frac{n_x}{N},~~
k_y=\frac{2\pi}{a}\frac{n_y}{N},~~
k_z=\frac{2\pi}{a}\frac{n_z}{N},
\end{equation}
where $n_x$, $n_y$ and $n_z$ run over all integer values in the range: 

\begin{equation}
-\frac{N}{2}<n_x,~n_y,~n_z\leq\frac{N}{2}.
\end{equation}
Here $N$ is called the size of the \textbf{k}-point mesh.

Both for even and odd mesh sizes the $\Gamma$-point is included in the $\Gamma$-centered Monkhorst-Pack scheme. The scheme is symmetric with respect to the $\Gamma$-point for odd mesh sizes. However it is asymmetric in case of even mesh sizes because $-N/2$~$n_x$, $-N/2$~$n_y$, and $-N/2$~$n_z$ coordinates are not included.

The second scheme is the original Monkhorst-Pack scheme, which is introduced for even meshes as:

\begin{equation}
k_x=\frac{2\pi}{a}\frac{n_x-\frac{1}{2}}{N},~~
k_y=\frac{2\pi}{a}\frac{n_y-\frac{1}{2}}{N},~~
k_z=\frac{2\pi}{a}\frac{n_z-\frac{1}{2}}{N},
\end{equation}
where $n_x$, $n_y$ and $n_z$ run over all integer values in the range: 

\begin{equation*}
-\frac{N}{2}<n_x,~n_y,~n_z\leq\frac{N}{2}.
\end{equation*}
Thus the scheme becomes symmetric with respect to the $\Gamma$-point, but the $\Gamma$-point itself is not included in the scheme.
The original Monkhorst-Pack scheme for odd mesh sizes coincides with the $\Gamma$-centered one.

Further the original Monkhorst-Pack scheme for even and odd meshes is designated as MP, and the $\Gamma$-centered scheme for even meshes is contracted as MP$\Gamma$. Both MP and MP$\Gamma$ schemes are described in the VASP manual \cite{vaspmanual}. In both schemes the weights of all \textbf{k}-points are set equal, with the unit value of the weights' sum.

\subsection{Reduction of the number of \textbf{k}-points}

The number of \textbf{k}-points in the Brillouin zone may be reduced due to symmetry reasons. This enables the calculation with the sufficient numbers of atoms and ionic configurations on the one hand, and with dense enough meshes in the Brillouin zone on the other hand. The procedure of the reduction of the number of \textbf{k}-points is close to that mentioned in paper \cite{Pozzo:PRB:2011}.

The \textbf{k}-points with the coordinates differing only by the sign are reduced to one. For example, \textbf{k}-points $(-0.4, -0.2, 0.2)$ and $(0.4, 0.2, 0.2)$ are merged into one \textbf{k}-point $(0.4, 0.2, 0.2)$ with the weight equal to the sum of weights of the initial \textbf{k}-points.

The \textbf{k}-points differing only by the order of coordinates are also reduced to one. For example, \textbf{k}-points $(0.4, 0.2, 0.2)$ and $(0.2, 0.4, 0.2)$ are merged into one \textbf{k}-point $(0.4, 0.2, 0.2)$ with the weight equal to the sum of weights of the initial \textbf{k}-points.

The absence of the particular direction during the simulation in the liquid phase may be mentioned as the support of the procedure described. However, in this paper the validity of this method is checked numerically, by the comparison of the results obtained for full and reduced meshes (see Fig.~\ref{Fig:KpointsDependence}). 

The reduction of the number of \textbf{k}-points is performed both for MP and MP$\Gamma$ meshes.

\subsection{Obtained results}

\begin{figure}
\includegraphics[width=0.95\columnwidth]{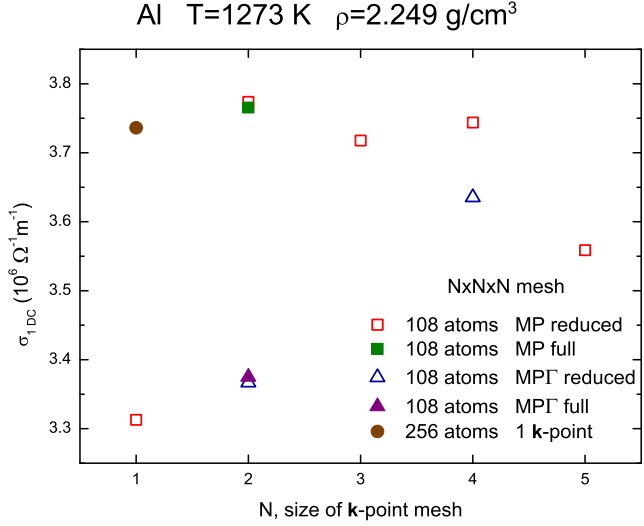}
\caption{(Color online) The dependence of static electrical conductivity on the number of \textbf{k}-points. Here $N$ is the size of the mesh. Filled square --- full MP scheme, empty squares --- reduced MP scheme. Filled triangle --- full MP$\Gamma$ scheme, empty triangles --- reduced MP$\Gamma$ scheme. The dependence on the number of \textbf{k}-points is investigated for 108 atoms in the supercell. Other simulation parameters are the same as in section \ref{Sec:Technique}. The result for the calculation with 256 atoms and $\Gamma$-point only is shown for comparison --- filled circle.}
\label{Fig:KpointsDependence}
\end{figure}

The obtained results on static electrical conductivity are presented in Fig.~\ref{Fig:KpointsDependence}.

Firstly the procedure of the reduction of \textbf{k}-points was validated. The results obtained both for full and reduced $2\times2\times2$ meshes are shown. The comparison is performed both for MP and MP$\Gamma$ schemes. The replacement of the full scheme by the reduced one yields less than 0.5\% change in static electrical conductivity. The changes in dynamic electrical conductivity are less than 2\% in the whole range of frequencies under consideration. This proves that the described procedure of reduction of the number of \textbf{k}-points is reliable enough and enables  investigation of the influence of the mesh size on results.

The difference between the results obtained using MP and MP$\Gamma$ schemes is shown for reduced $2\times2\times2$ and $4\times4\times4$ meshes. The changes are noticeable for a small mesh size and become less as the mesh size $N$ grows.

The values of static electrical conductivity obtained for all \textbf{k}-point meshes are less than 14\% larger than the value obtained for the $\Gamma$-point only. This value is close to the effect of the replacement of 108-atom supercell by one with 256 atoms. The result for 256 atoms and $\Gamma$-point only is also presented in Fig.~\ref{Fig:KpointsDependence}.

\begin{figure}
\includegraphics[width=0.95\columnwidth]{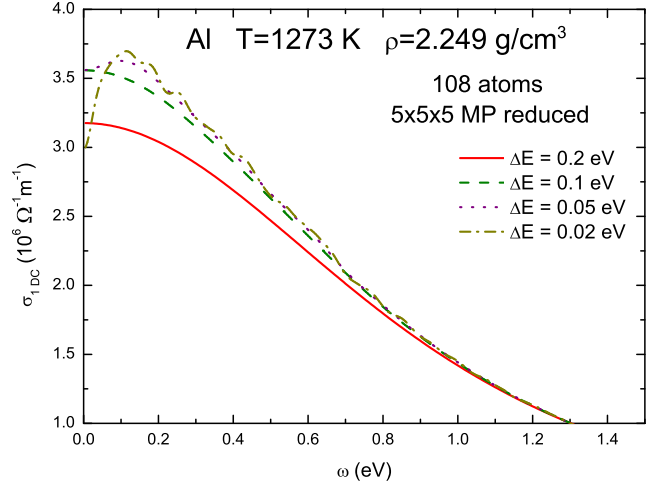}
\caption{(Color online) The dependence of the dynamic electrical conductivity on the broadening $\Delta E$ of the $\delta$-function. The number of atoms in the supercell is 108, the reduced MP $5\times5\times5$ mesh of \textbf{k}-points is used. Other simulation parameters are the same as described in section \ref{Sec:Technique}.}
\label{Fig:DeltaedependenceKpoints}
\end{figure}

However, the increment of the number of \textbf{k}-points hardly improves the convergence with the broadening $\Delta E$ of the $\delta$-function, described in subsection \ref{Subsec:Technique:Broadening}. The dependence of the obtained dynamic electrical conductivity on the broadening of the $\delta$-function at low frequencies is shown in Fig.~\ref{Fig:DeltaedependenceKpoints}. The results obtained for 108 atoms and $5\times5\times5$ reduced MP mesh (Fig.~\ref{Fig:DeltaedependenceKpoints}) are close to those obtained for 108 atoms and $\Gamma$-point only (Fig.~\ref{Fig:Deltaedependence}b). In both cases when the broadening $\Delta E$ is decreased the oscillations appear earlier than the convergence of dynamic electrical conductivity is achieved. Compared to this, increment of the number of atoms up to 256 (Fig.~\ref{Fig:Deltaedependence}a) leads to more significant improvements of the convergence with the broadening $\Delta E$. It is consistent with the statement of paper \cite{Lambert:PhysPlasmas:2011}, that the convergence with the broadening of the $\delta$-function may not be improved by the increment of the number of \textbf{k}-points.

\bibliographystyle{model1a-num-names}







\end{document}